\documentclass[aip,jcp,12pt,onecolumn,tightenlines,superscriptaddress]{revtex4}
\usepackage{amsfonts}
\usepackage{amsmath}
\usepackage{amssymb}
\usepackage{version}
\usepackage{color}
\usepackage{pgfplots}
\usepackage[capitalize]{cleveref}
\usepackage{graphicx}%
\includeversion{New_connection}
\excludeversion{Old_connection}

\usepgfplotslibrary{units}
\begin{document}
\title{The Challenge of Stochastic St{\o}rmer-Verlet Thermostats Generating Correct Statistics}

\author{Joshua Finkelstein}
\thanks{Present address: Theoretical Division, Los Alamos National Laboratory, Los Alamos, NM 87545, USA}
\affiliation{Department of Mathematics, Temple University, Philadelphia, PA 19122, USA}
  
\author{Chungho Cheng}
\affiliation{Department of Mechanical \& Aerospace Engineering,
  University of California, Davis, CA 95616, USA}
  
\author{Giacomo Fiorin}
\affiliation{Institute for Computational Molecular Science, Temple University,
                                 Philadelphia, PA 19122, USA}
\affiliation{National Heart, Lung and Blood Institute, Bethesda, MD 20892, USA}

\author{Benjamin Seibold}
\affiliation{Department of Mathematics, Temple University, Philadelphia, PA 19122, USA}
  
\author{Niels Gr{\o}nbech-Jensen}
\thanks{Corresponding author}
\email{ngjensen@math.ucdavis.edu}
\affiliation{Department of Mechanical \& Aerospace Engineering, University of California, Davis, CA 95616, USA}
\affiliation{Department of Mathematics, University of California, Davis, CA 95616, USA}

\today

$\;$\\

\begin{abstract}

\vspace{0.2 in}

\begin{center}
{\large Abstract}
\end{center}

\noindent
In light of the recently developed complete GJ set of single random variable stochastic, discrete-time St{\o}rmer-Verlet algorithms for statistically accurate simulations of Langevin equations, we investigate two outstanding questions: 1) Are there any algorithmic or statistical benefits from including multiple random variables per time-step, and 2) are there objective reasons for using one or more methods from the available set of statistically correct algorithms? To address the first question, we assume a general form for the discrete-time equations with two random variables and then follow the systematic, brute-force GJ methodology by enforcing correct thermodynamics in linear systems. It is concluded that correct configurational Boltzmann sampling of a particle in a harmonic potential implies correct configurational free-particle diffusion, and that these requirements only can be accomplished if the two random variables per time step are identical. We consequently submit that the GJ set represents all possible stochastic St{\o}rmer-Verlet methods that can reproduce time-step-independent statistics of linear systems. The second question is thus addressed within the GJ set. Based in part on numerical simulations of complex molecular systems, and in part on analytic scaling of time, we analyze the apparent difference in stability between different methods. We attribute this difference to the inherent time scaling in each method, and suggest that this scaling may lead to inconsistencies in the interpretation of dynamical and statistical simulation results. We therefore suggest that the method with the least inherent time-scaling, the GJ-I/GJF-2GJ method, be preferred for statistical applications where spurious rescaling of time is undesirable.
\end{abstract}
\maketitle

\section{Introduction}
\label{sec:intro}
Over the past several decades, discrete-time Langevin and Molecular Dynamics (MD) simulations have provided a wealth of information about the properties of nonlinear and complex systems \cite{AllenTildesley,Frenkel,Rapaport,Hoover_book,Leach}. While the simulations are intended to represent the continuous-time equations of motion, the inevitable temporal discretization alters not only the accuracy of the simulated trajectories, but in some cases also fundamental aspects of the system itself. Thus, an integral part of any simulation task is to explore and optimize the balance between the two conflicting objectives; namely simulation efficiency by increasing the discrete time step, and simulation accuracy by decreasing the discrete time step. For computational statistical mechanics, one of the key equations of motion to investigate is the Langevin equation \cite{Langevin,Langevin_Eq}, which is the topic we are concerned with in this paper,
\begin{eqnarray}
m\dot{v}+\alpha\dot{r} & = & f+\beta \; . \label{eq:Langevin}
\end{eqnarray}
Here $m$ is the mass of an object with spatial (configurational) coordinate $r$ and velocity $v=\dot{r}$. The object is subjected to a force $f$ and linear friction, which is represented by the non-negative constant $\alpha$. The fluctuation-dissipation relationship specifies that the thermal fluctuations $\beta$ can be represented through a temporally uncorrelated Gaussian variable \cite{Parisi}
\begin{subequations}
\begin{eqnarray}
\beta(t) & = & \sqrt{2\alpha\,k_BT}\,\sigma(t)\\
\langle\sigma(t)\rangle & = & 0 \\
\langle\sigma(t)\sigma(t^\prime)\rangle & = & \delta(t-t^\prime) \; , 
\end{eqnarray}\label{eq:Noise}
\end{subequations}
where $\delta(t)$ is Dirac's delta function, $k_B$ is Boltzmann's constant, and $T$ is the thermodynamic temperature.

Many methods (thermostats) for controlling the temperature of a simulated system have been developed, and most of them fall into two major categories: Deterministic (e.g., Nos{\'e}-Hoover \cite{Nose,Hoover,Hoover2,Martyna_92,Watanabe}) and stochastic (Langevin) thermostats (see the large body of work represented in, e.g., Refs.~\cite{SS,vgb_1982,Gunsteren,BBK,Skeel_2002,rc_2003,Melchionna_2007,Loncharich_2004,thalmann_2007,Mishra_1996,vdSpoel_2005,Vanden,Goga_2012,Paquet}). It has also been proposed to use hybrids of those two approaches in which the individual degree of freedom is largely deterministic, but with input from a global measure of kinetic temperature, which is stochastically regulated \cite{Bussi_2007,Bussi_07}. A common difficulty for methods that rely on a kinetic measure of temperature is that discrete time distorts the conjugated relationship between a coordinate and its momentum (velocity) (see, e.g., Refs.~\cite{Venneri,holian95,2GJ,GJ} for discussions). Thus, adjusting the kinetic temperature to the desired value will therefore inevitably lead to incorrect configurational sampling, which is typically the statistics of interest. The discrete-time inconsistencies between spatial and temporal coordinates have also caused similar imperfections in many stochastic thermostats as these methods have often relied on kinetic measures to assess thermal properties of a system. The root of the coordinate-velocity problem is in the introduction of approximate velocity variables to the original conservative St{\o}rmer-Verlet finite difference equation \cite{Stormer_1921,Verlet}, which does not have any explicit representation of velocity. The introduction of approximate finite difference velocities created the conservative velocity-explicit \cite{Swope,Beeman} and the leap-frog \cite{Buneman,Hockney} methods, which had the advantage of explicit inclusion of velocity, but without consistency of the conjugated relationship between the coordinate and its approximate velocity (see Ref.~\cite{GJ} for a discussion).

Focusing directly on configurational sampling in convex potentials, Leimkuhler and Matthews \cite{LM} developed the BAOAB method, which to our knowledge is the first to correctly sample Boltzmann statistics in a harmonic potential. Soon after, the stochastic GJF algorithm \cite{GJF1} demonstrated correct, time-step-independent configurational statistics for linear systems; i.e., correct drift, configurational diffusion as measured by the Einstein definition, and the Boltzmann distribution for a harmonic potential. Due to the above-mentioned discrete-time inconsistencies, it was shown that the accompanying on-site velocity coordinate exhibits statistical deviations that increase quadratically with the time step, thereby depressing kinetic measures in simulations with correct configurational sampling. By introducing a time scale revision to the BAOAB method, a very similar algorithm followed \cite{Sivak} with seemingly similar properties to the GJF method. A thermostat that can give correct thermodynamic response for both configurational and kinetic core quantities was demonstrated by the GJF-2GJ method \cite{2GJ}, which adopted the configurational sampling of the GJF method with a tailored half-step velocity definition that responds with correct, time-step-independent Maxwell-Boltzmann distribution for an object in a harmonic potential. It was further demonstrated that the diffusion measured by the velocity autocorrelation function also is correct if the appropriate Riemann sum is chosen for approximating the Green-Kubo integral of the velocity autocorrelation. The GJF-2GJ method is now available within the LAMMPS suite \cite{Plimpton,LAMMPS-Manual}. This method was recently generalized by the complete GJ set of stochastic, Verlet-type methods \cite{GJ}. These methods all display correct, time-step-independent configurational statistics, and they have accompanying half-step velocities that also respond with correct kinetic statistics. However, Ref.~\cite{GJ} illuminates that it is {\it not} possible to find an on-site velocity that correctly samples the Maxwell-Boltzmann distribution. Common for the GJ methods is that they include a single stochastic number for each degree of freedom in each time step. This is not unreasonable, since the Langevin equation (\ref{eq:Langevin}) has a single random sequence over a time step. However, direct integration of the Langevin equation over one time step reveals that this single sequence of randomness translates into two different, correlated noise terms in the discrete-time equations for the coordinate and its velocity \cite{vgb_1982} (see also Appendices \ref{appndxA} and \ref{appndxB} in this paper). Thus, it is obvious to methodically explore what benefits, if any, the inclusion of two random variables per time step can add to the quality of statistical simulations, and if it is possible to go beyond the quality that is exhibited by the GJ methods. Once it is understood which methods may be available for statistically robust sampling in discrete time, we further wish to determine if one can rationally discriminate among the available methods in order to select one or more methods with particularly desirable features. The objective in constructing methods that are statistically robust against the applied time step revolves around linear analysis and properties. We notice that for nonlinear systems, which are the typical systems for which one would apply the methods, additional discrete-time complexity enters into the dynamics \cite{2GJ2}. Further, as these new methods offer the possibility for applying large time steps, certain important quantities may unexpectedly suffer as a result. Thus, it is important to validate and investigate any developed method against systems that represent the true complexity and nonlinearity that the method needs to handle properly.
A review of the statistical accuracy of the methods in Refs.~\cite{BBK,Pastor_88,2GJ} was given in Ref.~\cite{2GJ}, and a comprehensive review of the configurational properties of the methods in Refs.~\cite{BBK,LM,GJF1} was recently given in Ref.~\cite{Josh}.

This presentation starts by reviewing the direct integration \cite{vgb_1982} approaches to the Langevin equation, and how that leads to a discretization with two random variables per time step. Following Ref.~\cite{GJ}, we then analyze a general stochastic form of the discrete-time St{\o}rmer-Verlet equations with two random variables with mutual correlation. We conclude that correct configurational statistics for linear systems can only be accomplished if the two random variables are fully correlated, implying that only a single random variable should be applied to simulations. This result points to the recently published complete set of GJ methods as representing all discrete-time options in the stochastic St{\o}rmer-Verlet form that can reliably provide robust statistics for relatively large time steps. We finally analyze the GJ methods for their mutual relationship and practical functionality. We observe a direct scaling relationship between the methods, and verify this scaling through comprehensive simulations of a complex molecular system in equilibrium.

\section{Method from Direct Integration}
\label{sec_lab_II}
The seemingly most direct approach to constructing discrete-time Langevin algorithms for simulations
is to analytically integrate the Langevin equation (\ref{eq:Langevin}) over one time step $\Delta{t}$,
from $t_n$ to $t_{n+1}=t_n+\Delta{t}$. This can be done in at least two ways, which
we have illustrated separately in Appendices \ref{appndxA} and \ref{appndxB}, respectively. Both of these approaches consist of exact integrals of the Langevin equation (\ref{eq:Langevin}) except for symmetric trapezoidal approximations to the integral over the conservative force $f$ in (\ref{eq:Langevin}). With $r^n\approx r(t_n)$ and $v^n\approx v(t_n)$, and $f^n$ representing the conservative force at time $t_n$, the immediate results of the two approaches can be expressed as follows:
\begin{description}
\item[A.] The result of approach A is given in Eqs.~(\ref{eq:Method_Ar}) and (\ref{eq:Method_Av})
\begin{subequations}
\begin{eqnarray}
r^{n+1} & = & r^n+\tilde{c}_3\Delta{t}\left[v^n+\frac{\Delta{t}}{2m}f^{n}\right]+\frac{\Delta{t}}{2m}\tilde{d}_r\beta_r^{n+1}\\
v^{n+1} & = & \tilde{c}_2v^n+\frac{\Delta{t}}{2m}(\tilde{c}_2f^n+f^{n+1})+\frac{1}{m}(\beta_{Av}^{n+1}-\frac{\alpha\Delta{t}}{2m}\tilde{d}_r\beta_r^{n+1})\; ,
\end{eqnarray}\label{eq:Method_A}
\end{subequations}
where $\tilde{c}_2$ and $\tilde{c}_3$ are given in Eqs.~(\ref{eq:tc2}) and (\ref{eq:tc3}), respectively. The noise terms $\tilde{d}_r\beta_r^{n+1}$ and $\beta_{Av}^{n+1}$ are given in Eqs.~(\ref{eq:drbeta_r}) and (\ref{eq:beta_Av}). Please see Appendix \ref{appndxA} for details on parameters and derivations.

\item[B.] The result of approach B is given in Eqs.~(\ref{eq:Method_Br}) and (\ref{eq:Method_Bv})
\begin{subequations}
\begin{eqnarray}
r^{n+1} & = & r^n+\tilde{c}_3\Delta{t}\left[v^n+\frac{\Delta{t}}{2m}f^{n}\right]+\frac{\Delta{t}}{2m}\tilde{d}_r\beta_r^{n+1}\\
v^{n+1} & = & \tilde{c}_2v^n+\frac{\Delta{t}}{2m}(\tilde{c}_2f^n+f^{n+1})
 +\frac{1}{m}\tilde{d}_{Bv}\beta_{Bv}^{n+1}\; ,
 \end{eqnarray}\label{eq:Method_B}
\end{subequations}
where the coefficient $\tilde{d}_{Bv}$ is given in Eq.~(\ref{eq:tdv}), and where the noise terms $\tilde{d}_r\beta_r^{n+1}$ and $\tilde{d}_{Bv}\beta_{Bv}^{n+1}$ are given in Eqs.~(\ref{eq:drbeta_r}) and (\ref{eq:tdBvbeta}).
Approach B coincides with that of Ref.~\cite{vgb_1982}. Please see Appendix \ref{appndxB} for details on parameters and derivations.
\end{description}
The two sets of Equations (\ref{eq:Method_1}) and (\ref{eq:Method_2}) (or, equivalently, Eqs.~(\ref{eq:Method_A}) and (\ref{eq:Method_B})) are identical except for the appearance of the noise term in the second equation of each set. However, by reviewing the definitions of the noise terms in Eqs.~(\ref{eq:beta_Av}),  (\ref{eq:drbeta_r}), and (\ref{eq:tdBvbeta}), we observe that
\begin{eqnarray}
\tilde{d}_{Bv}\beta_{Bv}^n & = & \beta_{Av}^n-\frac{\alpha\Delta{t}}{2m}\tilde{d}_r\beta_{r}^n\, .\label{eq:AsameB}
\end{eqnarray}
Thus, the two sets of Equations, describing methods A and B, are identical.
Curiously, implementation of this direct integration method (we will refer to the direct integration method in the form given in Appendix \ref{appndxB}) shows that it does not reproduce time-step-independent statistics if the conservative force represents a potential with non-zero curvature. However, the method and its derivation, which is correct in its fluctuation-dissipation relationship, allude to the rationale for considering the inclusion of two different noise terms for each time step, as these two noise terms represent different convolutions, one in velocity and one in position, of the integrated Langevin noise over the time step $\Delta{t}$. Thus, the following analysis generalizes the functional forms of these equations in order to investigate the possibilities for creating algorithms that can take advantage of two different noise terms per time step. This analysis will illuminate why the direct integration approach fails, and which possibilities for useful algorithms exist.

\section{General Discrete-Time Expressions}
\label{sec:general}
Inspired by the two direct integrations, A and B, outlined above, we write the general expressions in the form offered by Appendix~\ref{appndxB} \cite{vgb_1982}
\begin{subequations}
\begin{eqnarray}
r^{n+1} & = & r^n+d_2\Delta{t}\,v^n+d_3\frac{\Delta{t}^2}{2m}f^n+\frac{\Delta{t}}{2m}c_6\beta_r^{n+1} \label{eq:general_r}\\
v^{n+1} & = & c_2\,v^n+\frac{\Delta{t}}{2m}(d_6f^n+d_7f^{n+1})+\frac{d_v}{m}\beta_v^{n+1}\; ,\label{eq:general_v}
\end{eqnarray}\label{eq:general_rv}
\end{subequations}
where $c_2$ is the pivotal one-time-step velocity attenuation factor, and the other functional parameters are to be determined.
The two stochastic terms are given by
\begin{subequations}
\begin{eqnarray}
\beta_r^n & = & \sqrt{2\alpha\,k_BT\,\Delta{t}}\,\sigma_r^n\\
\beta_v^n & = & \sqrt{2\alpha\,k_BT\,\Delta{t}}\,\sigma_v^n\; , 
\end{eqnarray}
\end{subequations}
where the Gaussian random variables $\sigma_r^n$, $\sigma_v^n$ are independent across different time steps, and for the same step possess a joint probability distribution with covariance so that
\begin{subequations}
\begin{eqnarray}
\langle\sigma_r^n\sigma_r^\ell\rangle & = & \langle\sigma_v^n\sigma_v^\ell\rangle \; = \; \delta_{n,\ell}\\
\langle\sigma_r^n\sigma_v^\ell\rangle & = & \zeta \, \delta_{n,\ell}\; ,\label{eq:zeta}
\end{eqnarray}
\end{subequations}
where $\delta_{n,\ell}$ is Kronecker's delta function. Note that $\zeta$ is the correlation between the two noise terms of the same time step, with $|\zeta|\le1$. The choice $|\zeta|=1$ corresponds to the complete set of GJ methods \cite{GJ}, which has a single, shared noise term per time step.
As promoted in Refs.~\cite{2GJ,GJ}, the discrete-time velocity is an inherently ambiguous quantity, which for any given configurational coordinate $r^n$ can be defined with different features in mind. Thus, in order to design and evaluate the statistics of a method, it is convenient to first express this method without the direct representation of velocity, thereby focus exclusively on configurational statistics before the possibility of correct kinetic statistics is evaluated. We therefore rewrite Eq.~(\ref{eq:general_rv}) in the St{\o}rmer form
\begin{eqnarray}
r^{n+1} & = & 2c_1r^n-c_2r^{n-1}+\frac{\Delta{t}^2}{m}(c_3^\prime f^{n-1}+c_3f^n)+\frac{\Delta{t}}{2m}(2c_4\beta_v^n-c_2c_6\beta_r^n+c_6\beta_r^{n+1})\; , \label{eq:general_rr}
\end{eqnarray}
where we recognize the one time-step attenuation factor $c_2$ from the GJ set of methods \cite{GJ}. The initial relationship between parameters are:
\begin{subequations}
\begin{eqnarray}
2c_1 & = & 1+c_2\label{eq:c1}\\
2c_3^\prime & = & d_2d_6-d_3c_2\label{eq:c3prime}\\
2c_3 & = & d_3+d_2d_7\\
2c_4 & = & d_2d_v \; . 
\end{eqnarray}
\end{subequations}
\begin{description}
\item[{\underline{For $T=0$}},] we write the velocity $v^n$  in its general, central difference form
\begin{eqnarray}
v^{n+1} & = & \frac{\gamma_1r^{n+2}+\gamma_2r^{n+1}+\gamma_3r^{n}}{2\Delta{t}}\; ,\label{eq:general_vv}
\end{eqnarray}
where $\gamma_1+\gamma_2+\gamma_3=0$.
Inserting Eq.~(\ref{eq:general_rr}) into Eq.~(\ref{eq:general_vv}) yields
\begin{eqnarray}
v^{n+1} & = & c_2\,v^n+\frac{\Delta{t}}{2m}(-\gamma_3c_3^\prime f^{n-1}+(\gamma_1c_3^\prime-\gamma_3c_3)f^n+\gamma_1c_3f^{n+1})\label{eq:general_vv_prime}\; .
\end{eqnarray}
Comparing this expression with Eq.~(\ref{eq:general_v}) implies that $\gamma_3c_3^\prime=0$. It is further reasonable to require that the method can correctly mimic drift for $T=0$ and $f^n=f={\rm const}$. For $f^n=f={\rm const}$ and $T=0$, we write the constant drift velocity $v_d=f/\alpha$
\begin{eqnarray}
v_d & = & \frac{r^{n+1}-r^n}{\Delta{t}} \; = \; \frac{r^n-r^{n-1}}{\Delta{t}} \; = \; \frac{c_3^\prime+c_3}{1-c_2}\frac{\Delta{t}}{m}f \; = \; \frac{f}{\alpha}\\
\Rightarrow \; \; c_3^\prime+c_3 & = & \frac{1-c_2}{\alpha\Delta{t}/m}\; . \label{eq:_c3}
\end{eqnarray}
Appendix \ref{appndxC} investigates the special case $\gamma_3=0$, which allows $c_3^\prime\neq0$, and concludes that $c_3^\prime\neq0$ cannot lead to a meaningful method that supports a correct and time-step-independent Boltzmann distribution. Moreover, in the case of an explicit velocity, it can be shown that $\lim_{\alpha \to 0} c_3^\prime = 0$ is a necessary condition for the conservative method to be symplectic. Thus, we will proceed with the condition $c_3^\prime=0$ and
\begin{eqnarray}
c_3 & = & \frac{1-c_2}{\alpha\Delta{t}/m}\; , \label{eq:c3}
\end{eqnarray}
in conjunction with the GJ set of methods \cite{GJ}.
\end{description}
Using these observations, we can hereby write the equation for the trajectory $r^n$ in Eq.~(\ref{eq:general_rr}) as
\begin{eqnarray}
r^{n+1} & = & 2c_1r^n-c_2r^{n-1}+c_3\frac{\Delta{t}^2}{m}f^n+\frac{\Delta{t}}{2m}(2c_4\beta_v^n-c_5\beta_r^n+c_6\beta_r^{n+1})\; ,  \label{eq:general_rr_2}
\end{eqnarray}
where $c_5=c_2c_6$. Given the one time-step attenuation factor, $c_2$, there remain three parameters ($c_4$, $c_6$, and $\zeta$) in order to determine the noise terms such that the method correctly represents basic thermodynamic measures. Notice that the GJ set of methods \cite{GJ} is the same as Eq.~(\ref{eq:general_rr_2}) for $c_4=c_1c_3$, $c_6=c_3$ and $\zeta=1$ (i.e., $\beta_r^n=\beta_v^n$), where $\zeta$ expresses the correlation given in Eq.~(\ref{eq:zeta}). The functional parameter $c_2$ can be chosen among decaying functions for $\alpha\Delta{t}/m>0$ and functions that satisfy 
$c_2 \rightarrow 1-\frac{\alpha\Delta{t}}{m}$ for $\alpha\Delta{t}/m\rightarrow0$ \cite{GJ}, and this is the parameter that distinguishes the GJ methods from each other (see Eq.~(\ref{eq:method_c2}) below for key examples). The resulting method can only be meaningful for $|c_2|<1$, since $c_2$ is the velocity attenuation factor over the time $\Delta{t}$.
The two core statistical measures that we initially wish to reproduce are diffusion, as defined by the configurational Einstein definition, for a flat potential ($f^n=0$), and the Boltzmann distribution for a harmonic potential ($f^n=-\kappa r^n$ with $\kappa>0$).

\subsection{Diffusion in a flat potential, $f^n=0$}
\label{sec:diffusion}
For $f^n=0$, we follow Refs.~\cite{2GJ,GJ} and define the velocity $v^{n+\frac{1}{2}}$
\begin{eqnarray}
v^{n+\frac{1}{2}} & = & \frac{r^{n+1}-r^n}{\Delta{t}}\label{eq:half-step_w}
\end{eqnarray}
such that
\begin{eqnarray}
r^n-r^0 & = & \Delta{t}\sum_{k=0}^{n-1}v^{n+\frac{1}{2}\; . }\label{eq:displacement}
\end{eqnarray}
Inserting the velocity Eq.~(\ref{eq:half-step_w})  into Eq.~(\ref{eq:general_rr_2}) reads
\begin{eqnarray}
v^{n+\frac{1}{2}} & = & c_2v^{n-\frac{1}{2}}+\frac{1}{2m}(2c_4\beta_v^n-c_5\beta_r^n+c_6\beta_r^{n+1})\\
& = & c_2^nv^\frac{1}{2}+\frac{1}{2m}\sum_{k=0}^{n-1}c_2^k(2c_4\beta_v^{n-k}-c_5\beta_r^{n-k}+c_6\beta_r^{n-k+1})\; , 
\end{eqnarray}
which, when inserted into Eq.~(\ref{eq:displacement}), produces the diffusive displacement
\begin{eqnarray}
r^n-r^0 & = &\frac{1-c_2^n}{1-c_2}\Delta{t}\,v^{\frac{1}{2}}+\frac{\Delta{t}}{2m}\left[2c_4\beta_v^1-c_5\beta_r^1+c_6\frac{1-c_2^n}{1-c_2}\beta_r^{n+1}\right]\nonumber \\
& + & \frac{\Delta{t}}{2m}\frac{1}{1-c_2}\sum_{q=2}^{n}\left[(2c_4\beta_v^q-c_5\beta_r^q)(1-c_2^q)+c_6\beta_r^q(1+c_2^{q-1})\right]\; . 
\end{eqnarray}
The configurational diffusion is thus
\begin{eqnarray}
D_E & = & \lim_{n\Delta{t}\rightarrow\infty}\frac{\langle(r^n-r^0)^2\rangle}{2\,n\Delta{t}}\\
& = & \frac{1}{4c_3^2}\underbrace{\left[4c_4^2+(c_6-c_5)^2+4c_4(c_6-c_5)\zeta\right]}_{4c_3^2}\,\frac{k_BT}{\alpha}\; . 
\end{eqnarray}
Thus, basic diffusion requires that the noise parameters satisfy the condition
\begin{eqnarray}
4c_3^2 & = & 4c_4^2+(c_6-c_5)^2+4c_4\zeta(c_6-c_5)\nonumber \\
& = & (2c_4-c_5+c_6)^2-4c_4(c_6-c_5)(1-\zeta)\; . \label{eq:condition_diff}
\end{eqnarray}
Notice that this condition is satisfied for the direct integration method \cite{vgb_1982}, described in Appendices \ref{appndxA} and \ref{appndxB}, and summarized in Eqs.~(\ref{eq:Method_A})-(\ref{eq:AsameB}).

\subsection{Boltzmann distribution in a harmonic potential, $f^n=-\kappa r^n$}
\label{sec:harmonic}
For $f^n=-\kappa r^n=-\Omega_0^2mr^n$, where $\Omega_0=\sqrt{\kappa/m}$ is the natural frequency of the harmonic oscillator with spring constant $\kappa>0$, we adopt the methodology of Ref.~\cite{GJ} to find the autocorrelation $\langle r^nr^n\rangle$ from the linearized version of Eq.~(\ref{eq:general_rr_2}). In order to simplify the visual impression of the expressions, we define
\begin{eqnarray}
{\cal N}^{n+1} & = & \frac{\Delta{t}}{2m}\left(2c_4\beta_v^n-c_5\beta_r^n+c_6\beta_r^{n+1}\right)\label{eq:Big_noise}
\end{eqnarray}
and
\begin{eqnarray}
X & = & 1-\frac{c_3}{c_1}\frac{\Omega_0^2\Delta{t}^2}{2}\; , 
\end{eqnarray}
such that the linearized Eq.~(\ref{eq:general_rr_2}) becomes
\begin{eqnarray}
r^{n+1} & = & 2c_1Xr^n-c_2r^{n-1}+{\cal N}^{n+1}\; . \label{eq:general_rr_lin}
\end{eqnarray}

Multiplying Eq.~(\ref{eq:general_rr_lin}) with, respectively, $r^{n-1}$, $r^n$, and $r^{n+1}$,  and then making the statistical averages of the resulting equations, we get
\begin{eqnarray}
\left(\begin{array}{ccc}
1 & -2c_1X & c_2 \\
0 & 2c_1 & -2c_1X \\
c_2 & -2c_1X & 1\end{array}\right)\left(\begin{array}{c}\langle r^{n-1}r^{n+1}\rangle \\ \langle r^nr^{n+1}\rangle \\ \langle r^nr^n\rangle\end{array}\right) & = & \left(\begin{array}{c}\langle r^{n-1}{\cal N}^{n+1}\rangle \\ \langle r^{n}{\cal N}^{n+1}\rangle \\ \langle r^{n+1}{\cal N}^{n+1}\rangle \end{array}\right)\; , 
\end{eqnarray}
which can also be written
\begin{eqnarray}
\left(\begin{array}{ccc}
1 & -2c_1X & c_2 \\
0 & 1 & -x \\
0 & -X & 1\end{array}\right)\left(\begin{array}{c}\langle r^{n-1}r^{n+1}\rangle \\ \langle r^nr^{n+1}\rangle \\ \langle r^nr^n\rangle\end{array}\right) & = & \left(\begin{array}{c}0 \\ \frac{1}{2c_1}\langle r^{n}{\cal N}^{n+1}\rangle \\ \frac{1}{2c_1(1-c_2)}\langle r^{n+1}{\cal N}^{n+1}\rangle \end{array}\right) \; .
\end{eqnarray}
The autocorrelation of immediate interest is then directly given by
\begin{eqnarray}
\langle r^nr^n\rangle & = & \frac{\langle r^{n+1}{\cal N}^{n+1}\rangle+(1-c_2)X\langle r^n{\cal N}^{n+1}\rangle}{2c_1(1-c_2)(1-X^2)} \; = \; \frac{k_BT}{\kappa}\; .\label{eq:rnrn_1}
\end{eqnarray}
With
\begin{eqnarray}
\langle r^n{\cal N}^{n+1}\rangle & = & \left(\frac{\Delta{t}}{2m}\right)^2c_6(2c_4\zeta-c_5) \, 2\alpha\,k_BT\,\Delta{t}\\
\langle r^{n+1}{\cal N}^{n+1}\rangle & = &  2c_1X\langle r^n{\cal N}^{n+1}\rangle\nonumber\\
& + & \left(\frac{\Delta{t}}{2m}\right)^2\left[4c_4^2-4c_4c_5\zeta+c_5^2+c_6^2\right]\,2\alpha\,k_BT\,\Delta{t}\; , 
\end{eqnarray}
Eq.~(\ref{eq:rnrn_1}) can be written
\begin{eqnarray}
\langle r^nr^n\rangle & = & \frac{1}{2c_3^2}\frac{\overbrace{4c_4(c_4-c_5\zeta)+c_5^2+c_6^2}^{2c_3^2}+X\overbrace{2(2c_4\zeta-c_5)c_6}^{2c_3^2}}{1+X}\,\frac{k_BT}{\kappa}\; . \label{eq:condition_ho}
\end{eqnarray}
Thus, we must require both of the following conditions to be satisfied:
\begin{subequations}
\begin{eqnarray}
2c_3^2 & = & 4c_4(c_4-c_5\zeta)+c_5^2+c_6^2 \label{eq:condition_ho1}\\
c_3^2 & = & (2c_4\zeta-c_5)c_6\; . \label{eq:condition_ho2}
\end{eqnarray}\label{eq:condition_hos}
\end{subequations}
We notice that adding twice Eq.~(\ref{eq:condition_ho2}) to Eq.~(\ref{eq:condition_ho1}) yields precisely Eq.~(\ref{eq:condition_diff}). Thus, the requirement Eq.~(\ref{eq:condition_diff}) for correct free particle diffusion is redundant with the set Eq.~(\ref{eq:condition_hos}) that enforces the correct Boltzmann distribution for the harmonic potential.

\subsection{Configurationally accurate thermostats}
\label{sec:cond}
Since the three conditions, given in Eqs.~(\ref{eq:condition_diff}) and (\ref{eq:condition_hos}), determining the three parameters of the stochastic terms in Eq.~(\ref{eq:general_rr_2}), are redundant such that, e.g., Eq.~(\ref{eq:condition_diff}) can be discarded, we investigate only the two remaining oscillator conditions in Eq.~(\ref{eq:condition_hos}), which can be rewritten
\begin{subequations}
\begin{eqnarray}
\frac{1-c_2}{2}\left(\frac{c_6}{c_3}\right)^2 +\frac{1+c_2}{2}\left(\frac{c_4}{c_1c_3}\right)^2& = &1\label{eq:c4c6}\\
\left(\frac{c_4}{c_1c_3}\right)\left(\frac{c_6}{c_3}\right) \, \zeta & = & \frac{1}{2c_1}\left(1+c_2\left(\frac{c_6}{c_3}\right)^2\right)\; . \label{eq:zetac4c6}
\end{eqnarray}\label{eq:conditions_ho_bar_c6c4}
\end{subequations}
We here recognize the single random variable GJ result ($c_6=c_3$, $c_4=c_1c_3$, $\zeta=1$) as a solution to the two equations. Defining the two variables, $z_6$ and $z_4$,
\begin{subequations}
\begin{eqnarray}
z_6^2 & = & \frac{1-c_2}{2}\left(\frac{c_6}{c_3}\right)^2\label{eq_z6}\\
z_4^2 & = & \frac{1+c_2}{2}\left(\frac{c_4}{c_1c_3}\right)^2\; , \label{eq_z4}
\end{eqnarray}
\end{subequations}
we write Eq.~(\ref{eq:conditions_ho_bar_c6c4}) in the simple form
\begin{subequations}
\begin{eqnarray}
z_4^2+z_6^2 & = & 1\label{eq:condition_ho_z4z6_circ}\\
z_4\zeta & = & \frac{1}{2}\sqrt{\frac{1-c_2}{1+c_2}}\frac{1}{z_6}+\frac{c_2}{2c_1}\sqrt{\frac{1+c_2}{1-c_2}}z_6\; . \label{eq:condition_ho_z4z6_asymp}
\end{eqnarray}\label{eq:condition_ho_z4z6}
\end{subequations}

\begin{figure}[t]
\centering
\scalebox{0.6}{\centering \includegraphics[trim={2.5cm 4.0cm 1.0cm 7.0cm},clip]{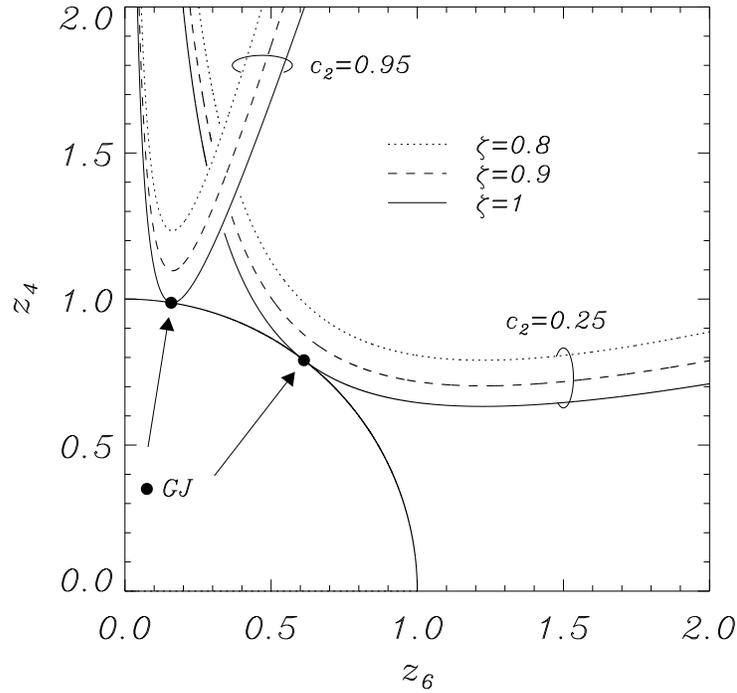}}
\caption{Sketch of the two conditions for time-step-independent statistics of the method. Eqs.~(\ref{eq:condition_ho_z4z6_circ}) (unit circle) and (\ref{eq:condition_ho_z4z6_asymp}). Two values of the one-time-step velocity attenuation factor $c_2$ are shown for the condition (\ref{eq:condition_ho_z4z6_asymp}), each for three different correlations $\zeta$ between the two random variables. Identified methods are labeled with a marker $\bullet$ at the intersection points of Eq.~(\ref{eq:intersection}) between the two conditions (\ref{eq:condition_ho_z4z6_circ}) and (\ref{eq:condition_ho_z4z6_asymp}). These are the GJ methods \cite{GJ}, $\zeta=1$, $c_4=c_1c_3$, and $c_6=c_3$.}
\label{fig_1}
\end{figure}
The task of identifying the solutions $(c_4,c_6,\zeta)$ now becomes one of finding the intersections between the two curves in Eq.~(\ref{eq:condition_ho_z4z6}). As illustrated in Fig.~\ref{fig_1} it is obvious that the curve described by Eq.~(\ref{eq:condition_ho_z4z6_circ}) is a unit circle, and that the curve described by Eq.~(\ref{eq:condition_ho_z4z6_asymp}) is outside the unit circle for both small and large positive values of $z_6$. Inserting Eq.~(\ref{eq:condition_ho_z4z6_asymp}) into Eq.~(\ref{eq:condition_ho_z4z6_circ}) gives
\begin{eqnarray}
(\zeta z_4)^2+z_6^2 -1 & = & -(1-\zeta^2)z_4^2\\
\Rightarrow \; \; \left(\frac{\sqrt{1-c_2}}{2}-\frac{z_6^2}{\sqrt{1-c_2}}\right)^2 & = & -2c_1(1-\zeta^2)z_4^2z_6^2\; . \label{eq:intersection}
\end{eqnarray}
While the left-hand side of this equation is non-negative, the right-hand side is non-positive. Thus, the intersection can {\it only} happen if
\begin{eqnarray}
\zeta^2 & = & 1 \; \; \Rightarrow \\
z_6^2 & = & \frac{1-c_2}{2} \; \; \Rightarrow \; \; z_4^2 \; = \; \frac{1+c_2}{2}\; ; 
\end{eqnarray}
i.e., {\it any} discrete-time method of the form Eq.~(\ref{eq:general_rv}), or equivalently Eq.~(\ref{eq:general_rr_2}), that is required to satisfy the basic thermodynamic requirements of Einstein diffusion and Boltzmann distribution in harmonic potentials must have noise correlation $|\zeta|=1$. Notice that the two cases $\zeta=\pm1$ simply imply the sign of $z_4z_6$, which is a distinction with no significance for the resulting algorithm. We therefore choose to set $\zeta=1$, which implies $z_4z_6>0$. It follows that {\bf there does not exist any method that can be expressed in the form Eq.~(\ref{eq:general_rv}) for which two different random variables can accomplish the required basic thermodynamic measures for the coordinate $r^n$}. The unique solution $\zeta=1$ is the one for which $c_6=c_3$ and $c_4=c_1c_3$. This unique possibility coincides with the complete GJ set of statistically sound methods \cite{GJ}, which has one random variable per time step.

\section{Properties of the GJ Set of Thermostats}
Having learned from the previous sections that no stochastic St{\o}rmer-Verlet-based thermostat can yield time-step-independent statistics with two different random variables per time step, we conclude that the complete GJ set of stochastic thermostats \cite{GJ}, using one stochastic variable per time step, is the only possible avenue for statistically accurate simulations of the Langevin equation for St{\o}rmer-Verlet type algorithms. This set is given by a free functional parameter, the one time step velocity attenuation factor $c_2$, which is a function of the reduced variable $\alpha\Delta{t}/m$. Using the result from Sec.~\ref{sec:cond}, the method can be written from Eq.~(\ref{eq:general_rr_2}) with $\zeta=1$ ($\beta_r^n=\beta_v^n=\beta^n$), $c_4=c_1c_3$, and $c_5=c_2c_6=c_2c_3$:
\begin{eqnarray}
r^{n+1} & = & 2c_1r^n-c_2r^{n-1}+c_3\frac{\Delta{t}^2}{m}f^n+\frac{c_3\Delta{t}}{2m}(\beta^n+\beta^{n+1})\; . \label{eq:GJ_sv}
\end{eqnarray}
The unique on-site $v^n$ and half-step $u^{n+\frac{1}{2}}$ velocities for this method were found in Ref.~\cite{GJ} to be
\begin{eqnarray}
v^n & = & \frac{r^{n+1}-(1-c_2)r^n-c_2r^{n-1}}{2\Delta{t}\sqrt{c_1c_3}}+\sqrt{\frac{c_3}{c_1}}\frac{1}{4m}(\beta^n-\beta^{n-1})\label{eq:on-site}\\
u^{n+\frac{1}{2}} & = & \frac{r^{n+1}-r^n}{\Delta{t}\sqrt{c_3}}\; , \label{eq:half-step}
\end{eqnarray}
respectively, where on-site and half-step velocities are defined \cite{GJ} such that
\begin{eqnarray}
\langle r^nv^n\rangle & = & 0\\
\langle(r^n+r^{n+1})u^{n+\frac{1}{2}}\rangle & = & 0\; , 
\end{eqnarray}
in accordance with the expectations from statistical mechanics \cite{Langevin_Eq}.
While it is not possible to find an on-site velocity that can provide correct, time-step-independent sampling of the kinetics, the unique half-step velocity was found to give
\begin{eqnarray}
\langle u^{n+\frac{1}{2}}u^{n+\frac{1}{2}}\rangle & = & \frac{k_BT}{m}
\end{eqnarray}
for the noisy harmonic oscillator, regardless of $\Delta{t}$ \cite{2GJ,GJ}. The combination of Eqs.~(\ref{eq:GJ_sv}) and (\ref{eq:on-site}) yields the specific velocity-explicit GJ method \cite{GJ}
\begin{subequations}
\begin{eqnarray}
r^{n+1} & = & r^n+\sqrt{c_1c_3}\,\Delta{t}\,v^n + \frac{c_3\Delta{t}^2}{2m}f^n+\frac{c_3\Delta{t}}{2m}\beta^{n+1}\label{eq:GJ_vv_r}\\
v^{n+1} & = & c_2v^n+\sqrt{\frac{c_3}{c_1}}\frac{\Delta{t}}{2m}(c_2f^n+f^{n+1})+\frac{\sqrt{c_1c_3}}{m}\beta^{n+1}\; . \label{eq:GJ_vv_v}
\end{eqnarray}\label{eq:GJ_vv}
\end{subequations}

This is the resulting form that determines the functional parameters in the general expression (\ref{eq:general_rv}) that we set out to investigate in this work. The resulting noise parameters that should be compared to the parameters $\tilde{d}_r$, $\tilde{d}_{Bv}$, and $\zeta_B$ for the direct integration method reviewed in Appendix \ref{appndxB} (and Fig.~\ref{fig_App_B}) are $d_r=c_3$, $d_v=\sqrt{c_1c_3}$, and $\zeta=1$. It is noticeable that these noise parameters are completely different from the ones arising from direct integration of the noise (see Appendix \ref{appndxB}). Further, while direct integration points to an exponential velocity attenuation factor $\tilde{c}_2$ in line with the GJ-II method, it also points to a parameter relationship ($d_2=d_3$, $d_6=c_2$, $d_7=1$ in Eq.~(\ref{eq:general_rv})), which is aligned with the GJ-I method where $c_1=c_3$ (see Eq.~(\ref{eq:method_c2}) below for relevant $c_2$ functions or Ref.~\cite{GJ} for details on the methods). Thus, the result of direct integration, leading to two different noise variables of each time step, is curiously inconsistent with basic statistical properties that are found in the GJ set.

Since the GJ methods are the only possibilities that will provide correct, time-step-independent statistics, we seek to determine if there are any of those methods that are especially advantageous. As indicated in Ref.~\cite{GJ}, the stability and harmonic period of these methods are the same if time and damping are scaled with the quantity $\sqrt{c_3/c_1}$ (see Eqs.~(56) and (70) in Ref.~\cite{GJ}). More generally, even beyond the linear limit, Eq.~(\ref{eq:GJ_vv}) can be written
\begin{subequations}
\begin{eqnarray}
r^{n+1} & = & r^n+c_1\tilde{\Delta{t}}\,v^n + c_1\frac{\tilde{\Delta{t}}^2}{2m}f^n+c_1\frac{\tilde{\Delta{t}}}{2m}\tilde{\beta}^{n+1}\label{eq:GJ_vv_r_scaled}\\
v^{n+1} & = & c_2v^n+\frac{\tilde{\Delta{t}}}{2m}(c_2f^n+f^{n+1})+c_1\frac{1}{m}\tilde{\beta}^{n+1}\; , \label{eq:GJ_vv_v_scaled}
\end{eqnarray}\label{eq:GJ_vv_scaled}
\end{subequations}
where
\begin{eqnarray}
\tilde{\beta}^n & = & \sqrt{2\tilde{\alpha}\,k_BT\,\tilde{\Delta{t}}} \, \sigma^n
\end{eqnarray}
and
\begin{eqnarray}
\tilde{\Delta{t}} & = & \sqrt{\frac{c_3}{c_1}}\,\Delta{t} \label{eq:scaling_Dt}\\
\tilde{\alpha} & = & \sqrt{\frac{c_3}{c_1}}\,\alpha\; . \label{eq:scaling_alpha}
\end{eqnarray}
\begin{figure}[t]
\centering
\scalebox{0.5}{\centering \includegraphics[trim={1cm 3cm 1.0cm 7.25cm},clip]{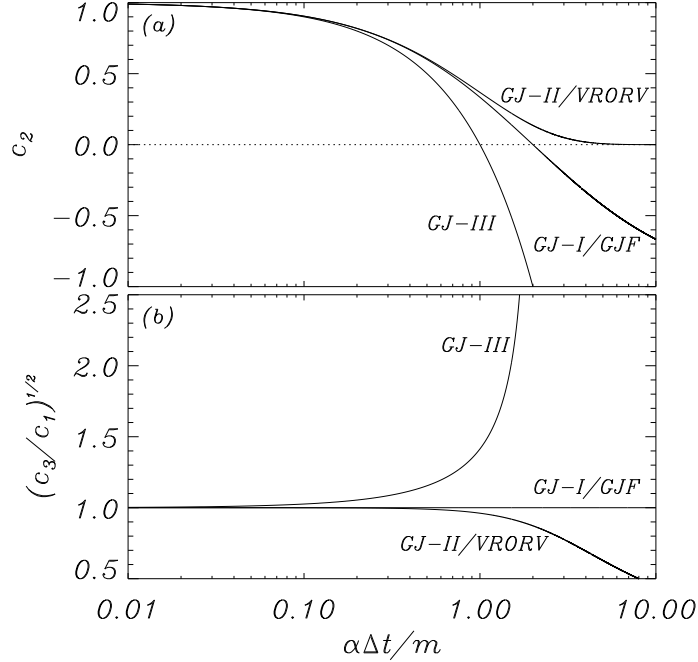}}
\caption{(a)~Single time step velocity attenuation factors for exemplified methods given in Eq.~(\ref{eq:method_c2}). (b)~Corresponding scaling of time and damping as given in Eqs.~(\ref{eq:scaling_Dt}) and (\ref{eq:scaling_alpha}), where $c_1$ and $c_3$ are given in Eqs.~(\ref{eq:c1}) and (\ref{eq:c3}).}
\label{fig:c2 plot}
\end{figure}
Consequently, the GJ set of methods can be mapped into a simpler form with revised parameters $\tilde{\Delta{t}}$ and $\tilde{\alpha}$. Three methods, GJ-I (GJF and GJ-I are the same for configurational sampling) \cite{GJF1,2GJ,GJ}, GJ-II (VRORV and GJ-II are the same for configurational sampling) \cite{Sivak,GJ}, and GJ-III \cite{GJ}, are exemplified in Fig.~\ref{fig:c2 plot}, where Fig.~\ref{fig:c2 plot}a shows the defining velocity attenuation factors $c_2$ defining the methods:
\begin{subequations}
\begin{eqnarray}
\text{GJ-I:} & c_2 & = \; \frac{1-\frac{\alpha\Delta{t}}{2m}}{1+\frac{\alpha\Delta{t}}{2m}} \\
\text{GJ-II:} & c_2 & = \; \exp(-\frac{\alpha\Delta{t}}{m}) \\
\text{GJ-III:} & c_2 & = \; 1-\frac{\alpha\Delta{t}}{m}\; .
\end{eqnarray}\label{eq:method_c2}
\end{subequations}
Figure~\ref{fig:c2 plot}b shows the corresponding scaling of time (or frequency \cite{GJ}). Notice that because $\alpha\Delta{t}/m$ appears explicitly in the parameter $c_2$, this mapping does not imply that the GJ set can be represented by a single method by scaling time and friction by the same factor. 
Nevertheless, consistent with the GJ method with the least time step distortion from the time step, the scaled method Eq.~(\ref{eq:GJ_vv_scaled}) is the GJ-I method \cite{GJ} (GJF-2GJ \cite{2GJ}) since this method is characterized by $c_1=c_3$. Thus, two GJ methods stand out as immediately useful. The first, for thermodynamic simulations, is the GJ-I (GJF-2GJ) method, which will provide correct thermodynamics in both configurational and kinetic sampling if one uses the half-step velocity for kinetics. This method exhibits the least amount of time scale distortion of all the methods. The second is the GJ-III method, which is characterized by $c_3=1$, such that not only kinetic statistics is accurate, but also drift and ballistic velocity measures from the half-step velocity Eq.~(\ref{eq:half-step}). This method does have time scale distortion, but an interest in drift and ballistic velocities would suggest an interest in dynamics, which in turn would necessitate rather small time steps for temporal accuracy. \\

\section{Numerical Results}
\begin{figure}
    \centering
\scalebox{0.6}{\centering \includegraphics[trim={1cm 6.0cm 1.0cm 7.0cm},clip]{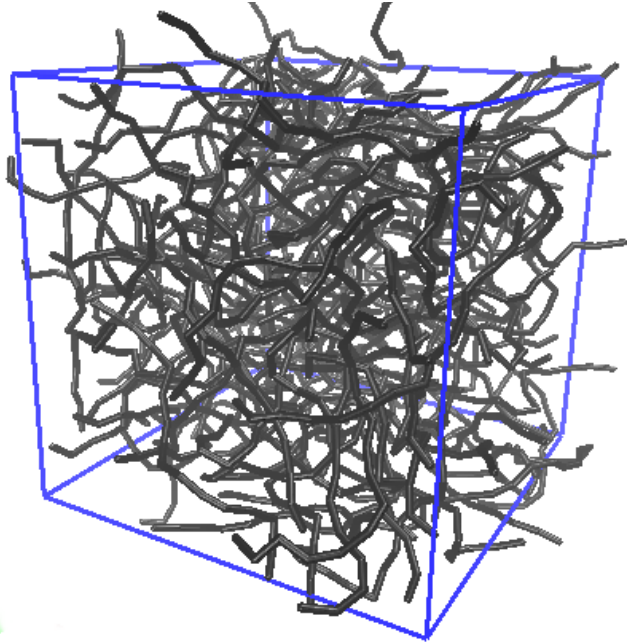}}
    \caption{Coarse-grained polyethylene system \cite{Josh} inside a periodic box with dimensions $58.065\text{~\AA} \times 58.065\text{~\AA} \times 58.065\text{~\AA}$. This figure was made using VMD \cite{vmd}.}
    \label{fig:cg_pe}
\end{figure}

Coarse-grained (CG) simulations of molecular systems often lead to much larger values of $\alpha \Delta t/m$ than those that would be seen in the corresponding atomistic simulations. This larger regime of $\alpha \Delta t /m$ is precisely the area of parameter space where the GJ methods differ significantly from each other (c.f.~\cref{fig:c2 plot}), and where they differ from other methods as well. For this reason, we sought to numerically study the select GJ methods, GJ-I, GJ-II and GJ-III, using a CG system. For each of the considered methods, a CG polyethylene melt consisting of 128 coarse-grained $\text{C}_{48} \text{H}_{98}$ hydrocarbon chains, as pictured in \cref{fig:cg_pe}, was simulated for 200 ns at 450 K after an initial NPT equilibration of 10 ns. Accordingly, we chose $\alpha/m =0.1$ fs$^{-1}$ (a large value, by atomistic standards) and used a wide range of choices for the simulation time step $\Delta t$, where this range differed for each method due to the method-dependent time scaling. To coarse-grain the polyethylene, the established Shinoda-DeVane-Klein (SDK) coarse-graining model in Ref.~\cite{SDK} was used, resulting in a new CG system with 2,048 CG particles and a CG potential energy consisting of bond, angle and Lennard-Jones force terms. Details on the actual force field parameters can be found in Refs.~\cite{Josh,SDK}. 

We studied both the translational diffusion rate and the rotational diffusion, or relaxation, time scale by computing the time auto-correlation function (ACF) of the end-to-end vector of each CG polymer chain:
\begin{eqnarray}
\langle P_2(\theta(\tau)) \rangle = \bigg \langle\frac{1}{2}(3\cos^2(\theta(\tau))-1) \bigg\rangle\; ,  \label{eq:P2}
\end{eqnarray}
where
\begin{align}
   \theta(\tau) = \arccos \bigg( \frac{\mathbf{R}(t) \cdot \mathbf{R}(t+\tau)}{|\mathbf{R}(t)| |\mathbf{R}(t+\tau)|}  \bigg)
\end{align}
is the angle between the end-to-end vector $\mathbf{R}$ of a given chain at time $t$ and time $t+\tau$. The brackets $\langle\cdot\rangle$ indicate averaging among the chains and over trajectory frames $t$.

\begin{figure}[t]
\centering
\scalebox{0.7}{\centering \includegraphics[trim={1cm 4.0cm 1.0cm 10.0cm},clip]{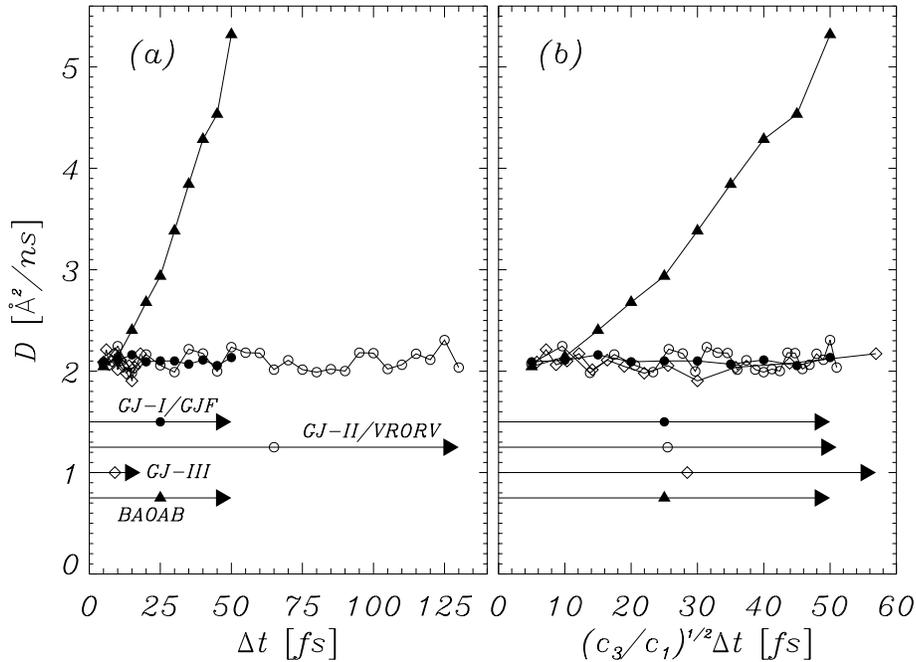}}
\caption{CG polyethylene diffusion coefficient $D$ as a function of time step $\Delta{t}$ 
for $\alpha/m =0.1$ fs$^{-1}$.  Horizontal arrows indicate the observed stability ranges of the different labeled simulation methods. (a)~Unscaled time step on horizontal axis. (b)~Same data as in (a), but with time step scaled by $\sqrt{c_3/c_1}$. Notice that neither GJ-I/GJF nor BAOAB scale time for high frequency motion as a function of $\alpha\Delta{t}/m$.}
\label{fig:Diffusion}
\end{figure}

\begin{figure}[t]
\centering
\scalebox{0.6}{\centering \includegraphics[trim={1cm 2.0cm 1.0cm 7.0cm},clip]{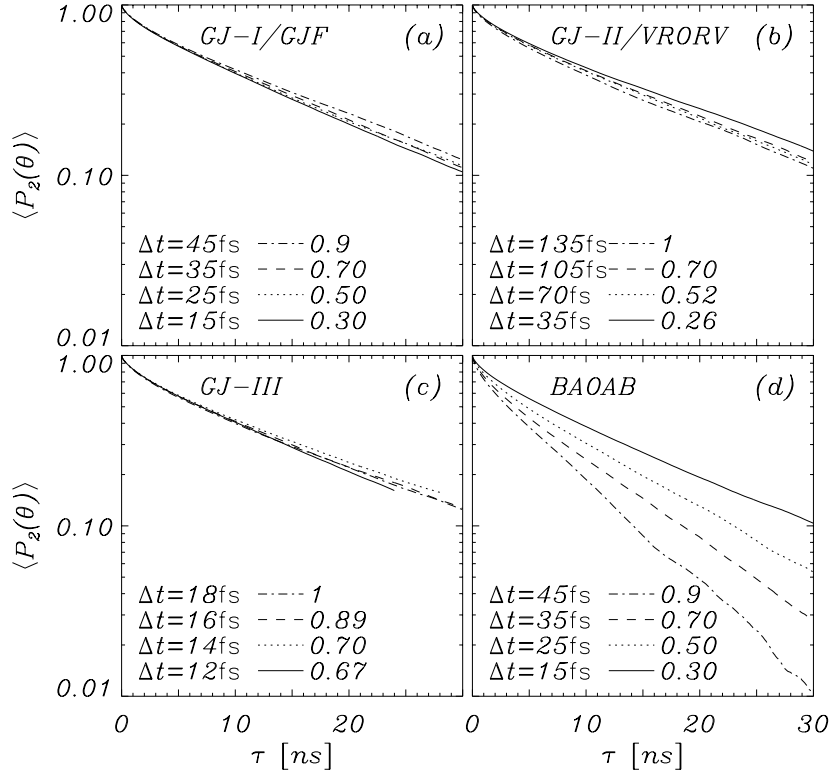}}
\caption{Temporal correlation $\langle P_2(\theta(\tau))\rangle$, indicating rotational diffusion, given in Eq.~(\ref{eq:P2}) for CG polyethylene with
$\alpha/m =0.1$ fs$^{-1}$. Results of different integration methods and different time steps $\Delta{t}$ across the relevant stability ranges are shown. The largest $\Delta{t}$ shown is at or near the numerically largest observed stable time step. Decimal number next to indicated $\Delta{t}$ is corresponding fraction of observed time step limit. (a)~GJ-I/GJF, (b)~GJ-II/VRORV, (c)~G-III, and (d)~BAOAB.
}
\label{fig:P2}
\end{figure}

\begin{figure}[t]
\centering
\scalebox{0.7}{\centering \includegraphics[trim={1cm 4.0cm 1.0cm 10.0cm},clip]{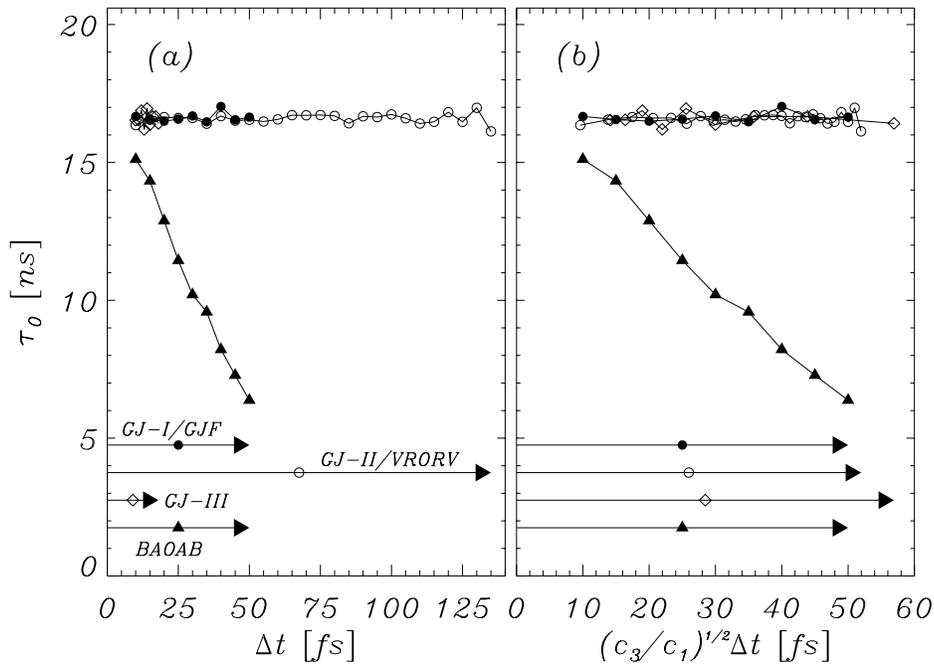}}
\caption{Long time characteristic decay time $\tau_0$ for $\langle P_2(\theta(\tau))\rangle\propto\exp(-\tau/\tau_0)$ ($10{\rm ns}\le\tau\)) as measured from data exemplified in Fig.~\ref{fig:P2}, for different integration methods and different time steps $\Delta{t}$ across the relevant stability ranges. (a)~Unscaled time step on horizontal axis. (b)~Same data as in (a), but with time step scaled by $\sqrt{c_3/c_1}$. Arrows indicate the range of time steps that are allowed for each method. Notice that neither GJ-I/GJF nor BAOAB scale time for high frequency motion as a function of $\alpha\Delta{t}/m$.
}
\label{fig:tau_0}
\end{figure}

In \cref{fig:Diffusion} we display the linear diffusion coefficient as a function of time step, both scaled and non-scaled. These data points are obtained from an average of 20 (GJ-II and GJ-III) and 100 (GJ-I and BAOAB) different simulations. In \cref{fig:P2} we present the temporal ACF of $P_2(\theta)$ for the different methods with a range of time steps. These curves are each generated as an average from 20 different simulations, and the resulting standard deviation of the averaged curves is estimated to be around $2\times10^{-3}$. In both figures, a comparison to the BAOAB method is made as a reference to the importance of the time scale revision that produced the VRORV method from BAOAB. Linear diffusion is measured by the configurational mean-squared distance over time. The diffusive properties observed in Figs.~\ref{fig:Diffusion}-\ref{fig:tau_0} are both time-step-independent and overall the same for the three studied GJ methods. The stability ranges, as indicated by the data and the horizontal arrows, are apparently vastly different. As seen in Fig.~\ref{fig:Diffusion}b, this difference in stability, however, is accurately compensated for by the above-mentioned time scaling factor $\sqrt{c_3/c_1}$, which scales the results of GJ-II and GJ-III almost into concert with those of GJ-I and BAOAB. This scaling is analytically shown in Ref.~\cite{GJ} to enter the stability range. The physical interpretation for the apparent differences between stability of the different methods is therefore mostly an illusion, since the scaling can equally well be interpreted as a change in oscillation frequency, as also shown in Ref.~\cite{GJ}. Thus, a method of this kind will always have a stability range in time step that is uniquely linked to the period of motion. It follows that the reason for, e.g., the GJ-II method to have a stability range significantly larger than, e.g., GJ-I, is that the GJ-II method has slowed down the dynamics, and the mobility per time step (relative to the stability limit) has therefore not actually changed. This is also the time scaling that links BAOAB and VRORV. As we obtain overall time step independent diffusion results, linear diffusion seems in these simulations to be dominated by the mechanism of particles in a viscous medium, which is a feature that the GJ methods are designed to reproduce regardless of the time scale of the dynamics.

For rotational diffusion, we observe in Figs.~\ref{fig:P2}a--c that the temporal $P_2$ correlation is seemingly independent of both time step and the specific GJ method used, even if different time steps can be applied to the different methods. In contrast, Fig.~\ref{fig:P2}d shows that the same simulation conducted with the BAOAB method \cite{LM}, which is not specifically designed to yield the correct free particle diffusivity, exhibits a visible time step dependence in this measure of diffusion. Note here that as the time step is increased, the rotational relaxation time scale decreases, that is, the system's dynamics speed up. This is consistent with earlier findings (also displayed in \cref{fig:Diffusion}) on the increased linear diffusivity of BAOAB observed for the regime of larger $\alpha \Delta t/m$ in Refs.~\cite{Josh,Sivak}. Emphasizing this observation, we notice that the correlation shown in Fig.~\ref{fig:P2} seems to be dominated by a single time scale for about $\tau>10{\rm ns}$. Extracting this decay rate for data by fitting $\langle P_2(\theta(\tau))\rangle$ to
\begin{eqnarray}
\langle P_2(\theta(\tau))\rangle & \propto & \exp(-\tau/\tau_0)
\end{eqnarray}
for $\tau>10{\rm ns}$,
we obtain Fig.~\ref{fig:tau_0} in which the characteristic correlation decay time $\tau_0$ is shown as a function of $\Delta{t}$ for the different methods reviewed in this work. As in Fig.~\ref{fig:Diffusion}, where linear diffusion is shown, we display the data for both unscaled and scaled time steps in order to illuminate the similarities between the methods. In concert with Fig.~\ref{fig:Diffusion}, the GJ methods exhibit nearly time step independent decay time of the rotational order across all three methods, while the BAOAB method shows that the orientational order is lost at an increasing rate as the time step is increased. This is consistent with the linear diffusion properties.

Having constructed the set of GJ methods to preserve the Einstein diffusion relation in the case of a free particle, we are not surprised that GJ-I/GJF, which is the GJ method that does not scale time as a function of $\alpha\Delta{t}/m$, exhibits translational diffusivity independent of $\alpha \Delta t/m$. This was also confirmed previously in Ref.~\cite{Josh}. However, given the complexity of the polyethylene system and the highly non-linear interactions between the chains, it is not obvious that the relaxation time scales of the slowest motions should remain unaffected by the implicit time scaling occurring with the GJ-II and GJ-III methods. Instead, \cref{fig:Diffusion,fig:P2,fig:tau_0} show that these relaxation times are controlled only by the friction parameter $\alpha/m$, indicating that the observed diffusion should be understood as described by an effective viscous medium.

\section{Discussion and Conclusion}
\label{sec:discussion}

We have conducted a thorough methodical investigation of the possibilities for creating discrete-time, stochastic St{\o}rmer-Verlet algorithms that correctly represent the equilibrium statistics of Langevin equations with conservative force fields regardless of the discrete time step. An inspection of the direct integration result, where only the integral over conservative force is approximated, suggests that a discrete-time algorithm may benefit from two different noise terms per time step, since the noise term in the continuous Langevin equation is integrated differently for the kinetic and configurational coordinates. Following the outline of the analysis that led to the GJ set of methods \cite{GJ}, we conduct a comprehensive analysis of a completely general form of a stochastic St{\o}rmer-Verlet expression that include different noise terms. This is investigated for its ability to yield numerical methods that respond with correct statistical measures of diffusion and Boltzmann distributions when simulating linear (harmonic) systems. The somewhat surprising, albeit definite, result is that this can {\it only} be accomplished if the two noise terms are identical. Consequently, the only methods of this kind that can accomplish the basic statistical objectives are the previously identified GJ set of methods, which includes a single noise term per time step \cite{GJ}, and to which discrete-time velocity variables that yield comparable correct kinetic statistics can be associated.

Given that recent developments of methods, starting with the BAOAB method \cite{LM}, have been succeeding in correctly describing the Boltzmann distribution for the configurational coordinate in convex potentials, we have conducted simulations to investigate the more subtle question of time scaling in diffusive behavior in dense, nonlinear systems, exemplified by a coarse-grained polyethylene system. Unlike the BAOAB method, the GJ set has been designed to provide the correct diffusion constant as derived from the configurational Einstein definition. However, this diffusion is for a free particle in a trivially flat energy landscape. The dense polyethylene system offers a complex and nonlinear environment in which the accuracy of diffusion of a free particle may not translate into a reliably calculated diffusion constant by such methods. The results for both linear and rotational diffusion show, however, that the three characteristic GJ methods reproduce the same time step independent diffusion constant throughout their entire stability ranges, leading to the conclusion that the designed free particle diffusion properties of the GJ methods translate into robust diffusion properties for more complex systems. In contrast, the BAOAB method exhibits significant deviations as the time step is increased for both linear and rotational diffusion, consistent with this method not having specifically enforced free particle diffusion in its design. It is noted that this discrepancy is purely in the time scale, as illuminated by the VRORV method \cite{Sivak}, which intersects in its configurational sampling with the GJ-II method. The VRORV method was derived by introducing a time scaling of the BAOAB method, specifically to address free particle diffusion, and the success of this revision is visible in Figs.~\ref{fig:Diffusion}-\ref{fig:tau_0}. The introduced time scaling of the methods have some important consequences that have been alluded to in Ref.~\cite{GJ}, and which can be seen in Fig.~\ref{fig:Diffusion} as an apparent difference in stability range of the different methods (the two extremes of the displayed methods being GJ-II and GJ-III). It is important to realize that these differences are not due to actual differences in stability, since this is solely linked to the time scaling $\sqrt{c_3/c_1}$, which simply affects the dynamics of the simulated behavior. For example, as described in Ref.~\cite{GJ}, a harmonic oscillator simulated by these methods will have a scaled frequency that ensures all these methods having the same stability limit as measured relative to the period of motion. It is therefore important that one does not infer sampling efficiency or stability directly from $\Delta{t}$, but considers the time scaling inherent to each method such that $\sqrt{c_3/c_1}\Delta{t}$ is the relevant time step for stability and sampling efficiency. In light of this discussion, we point to BAOAB and GJ-I/GJF, since neither of these are subject to the above-mentioned time scaling. Given that GJ-I/GJF-2GJ is the GJ method with the least time distortion ($\sqrt{c_3/c_1}=1$), we suggest that this is the method that generally provides the most reliable simulation results, especially for applications in which spurious time rescaling can cause difficulties with the interpretation of simulation data that are obtained using the large time steps offered by these methods. As mentioned earlier in this paper, the GJ-I/GJF-2GJ method \cite{2GJ} is available for use in the LAMMPS simulation suite \cite{Plimpton,LAMMPS-Manual}.

\section{Acknowledgements}
This research includes calculations carried out on HPC resources supported in part by the National Science Foundation through major research instrumentation grant number 1625061 and by the US Army Research Laboratory under contract number W911NF-16-2-0189, the latter also supporting the work of GF. Work of BS has been supported by National Science Foundation grant DMS--1719640.
\section{Data Availability Statement}
The data that support the findings of this study are available from the corresponding author upon reasonable request.


\appendix
\section{Direct Integration, Method A}
\label{appndxA}
We here start with the direct integration of the Langevin equation (\ref{eq:Langevin})
\begin{eqnarray}
m\int_{t_n}^t\dot{v}\,ds + \alpha\int_{t_n}^t\dot{r}\,ds& = & \int_{t_n}^t(f+\beta)\,ds\; ,
\end{eqnarray}
where $\beta(s)$ is given by Eq.~(\ref{eq:Noise}).
This immediately becomes
\begin{eqnarray}
v(t)+\frac{\alpha}{m}r(t) & = & v^n+\frac{\alpha}{m}r^n+\frac{1}{m}\int_{t_n}^t(f+\beta)\,ds\; , \label{eq:v_of_t_1}
\end{eqnarray}
which for $t=t_{n+1}=t_n+\Delta{t}$ can be written
\begin{eqnarray}
v(t_{n+1}) & = & v^n-\frac{\alpha}{m}(r(t_{n+1})-r^n)
+\frac{1}{m}\int_{t_n}^{t_{n+1}}f(s)\,ds+\frac{1}{m}\beta_{Av}^{n+1} \label{eq:v_of_tn_1}
\end{eqnarray}
with
\begin{eqnarray}
\beta_{Av}^{n+1} & = & \int_{t_n}^{t_{n+1}}\beta(s)\,ds \label{eq:beta_Av}
\end{eqnarray}
such that  $\langle\beta_{Av}^n\rangle=0$ and
\begin{eqnarray}
\langle\beta_{Av}^n\beta_{Av}^l\rangle & = & 2\alpha\,k_BT\,\Delta{t}\,\delta_{n,\ell} \; .
\end{eqnarray}
Equation (\ref{eq:v_of_t_1}) can be written
\begin{eqnarray}
\frac{d(\mu r)}{dt} & = & (v^n+\frac{\alpha}{m}r^n)\mu+\frac{\mu}{m}\int_{t_n}^t(f+\beta)\,ds\; , 
\end{eqnarray}
where $\mu(t)=\exp(\frac{\alpha}{m}(t-t_n))$ is the beneficial choice. Integration yields the coordinate $r(t)$
\begin{eqnarray}
\mu(t)r(t) & = & r^n+\frac{m}{\alpha}(v^n+\frac{\alpha}{m}r^n)(e^{\frac{\alpha}{m}(t-t_n)}-1)\\
&&+\int_{t_n}^t\frac{\mu(t^\prime)}{m}\int_{t_n}^{t^\prime}(f(s)+\beta(s))\,ds\,dt^\prime\; , \nonumber
\end{eqnarray}
which can be rewritten
\begin{eqnarray}
r(t) & = & r^n+\frac{1-e^{-\frac{\alpha}{m}(t-t_n)}}{(t-t_n)\alpha/m}(t-t_n)\,v^n +  \frac{t-t_n}{m}\int_{t_n}^tf(s)\frac{1-e^{-\frac{\alpha}{m}(t-s)}}{(t-t_n)\alpha/m}\,ds
\nonumber \\ &&
+ \frac{t-t_n}{2m}\int_{t_n}^t2\beta(s)\frac{1-e^{-\frac{\alpha}{m}(t-s)}}{(t-t_n)\alpha/m}\,ds \; . \label{eq:r_of_t_1}
\end{eqnarray}
For $t=t_{n+1}$ the equation reads
\begin{eqnarray}
r(t_{n+1}) & = & r^n+\tilde{c}_3\Delta{t}\,v^n +  \frac{\Delta{t}}{m}\int_{t_n}^{t_{n+1}}f(s)\frac{1-e^{-\frac{\alpha}{m}(t_{n+1}-s)}}{\alpha\Delta{t}/m}\,ds
+  \frac{\Delta{t}}{2m}\tilde{d}_r\beta_r^{n+1}\label{eq:r_of_tn_1}
\end{eqnarray}
with
\begin{eqnarray}
\tilde{c}_2 & = & e^{-\frac{\alpha\Delta{t}}{m}} \label{eq:tc2}\\
\tilde{c}_3 & = & \frac{1-\tilde{c}_2}{\frac{\alpha\Delta{t}}{m}}\label{eq:tc3}\\
\tilde{d}_r\beta_r^{n+1} & = & \int_{t_n}^{t_{n+1}}2\beta(s)\frac{1-e^{-\frac{\alpha}{m}(t_{n+1}-s)}}{\alpha\Delta{t}/m}\,ds\; , \label{eq:drbeta_r}
\end{eqnarray}
where $\beta_r^{n+1}$ is a Gaussian random number with $\langle\beta_r^{n}\rangle=0$ and
\begin{eqnarray}
&&\langle\beta_r^n\beta_r^l\rangle \; = \; 
2\alpha\,k_BT\,\Delta{t}\,\delta_{n,\ell} \; .
\label{eq:brbr_1} \end{eqnarray}
The coefficient $\tilde{d}_r$ is given by
\begin{eqnarray}
\tilde{d}_r^2 & = &  \frac{1 -2\tilde{c}_3+\tilde{c}_1\tilde{c}_3}{\left(\frac{\alpha\Delta{t}}{2m}\right)^{2}}\label{eq:tdr}\\
2\tilde{c}_1 & = & 1+\tilde{c}_2\; . \label{eq:tc1}
\end{eqnarray}

\begin{figure}[t]
\centering
\scalebox{0.5}{\centering \includegraphics[trim={2.5cm 4.0cm 1.0cm 8.0cm},clip]{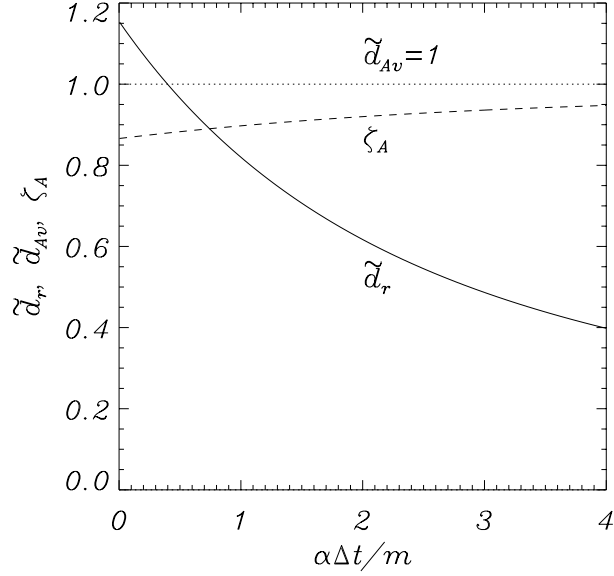}}
\caption{Noise parameters for the direct integration approach of Appendix \ref{appndxA} as a function of reduced time step $\alpha\Delta{t}/m$. Displayed parameters $\tilde{d}_r$ and $\zeta_A$ are from Eqs.~(\ref{eq:tdr}) and (\ref{eq:brbv_1}), respectively. We define $\tilde{d}_{Av}=1$ from Eq.~(\ref{eq:v_of_tn_1}) since the coefficient to $\beta_{Av}^{n+1}$ is $\frac{1}{m}$.}
\label{fig_App_A}
\end{figure}

The resulting numerical method appears by making the appropriate symmetric trapezoidal approximations to the integrals over the conservative $f$ in Eqs.~(\ref{eq:r_of_tn_1}) and (\ref{eq:v_of_tn_1}):
\begin{subequations}
\begin{eqnarray}
r^{n+1} & = & r^n+\tilde{c}_3\Delta{t}\left[v^n+\frac{\Delta{t}}{2m}f^{n}\right]+\frac{\Delta{t}}{2m}\tilde{d}_r\beta_r^{n+1}\label{eq:Method_Ar}\\
v^{n+1} & = & v^n-\frac{\alpha}{m}(r^{n+1}-r^n)+\frac{\Delta{t}}{2m}(f^n+f^{n+1})+\frac{1}{m}\beta_{Av}^{n+1} \\
& = & \tilde{c}_2v^n+\frac{\Delta{t}}{2m}(\tilde{c}_2f^n+f^{n+1})+\frac{1}{m}(\beta_{Av}^{n+1}-\frac{\alpha\Delta{t}}{2m}\tilde{d}_r\beta_r^{n+1})\; , \label{eq:Method_Av}
\end{eqnarray}\label{eq:Method_1}
\end{subequations}
where the noise correlation between $\beta_{Av}^n$ and $\beta_r^l$ is
\begin{eqnarray}
\zeta_A & = & \frac{\langle\beta_{Av}^n\beta_r^l\rangle}{\sqrt{\langle\beta_r^n\beta_r^n\rangle\,\langle\beta_{Av}^n\beta_{Av}^n\rangle}} \; = \; \frac{1-\tilde{c}_3}{\sqrt{1-2\tilde{c}_3+\tilde{c}_1\tilde{c}_3}}\,  \delta_{n,l}\; .  \label{eq:brbv_1}
\end{eqnarray}
Please see Fig.~\ref{fig_App_A} for the noise parameters $\tilde{d}_r$ and $\zeta_A$. Equation (\ref{eq:Method_1}) can finally be written
\begin{eqnarray}
r^{n+1} & = & 2\tilde{c}_1r^n-\tilde{c}_2r^{n-1}+\frac{\tilde{c}_3\Delta{t}^2}{m}f^n
+\frac{\Delta{t}}{2m}\left(2\tilde{c}_3\beta_{Av}^n-\tilde{d}_r\beta_r^n+\tilde{d}_r\beta_r^{n+1}\right)\; .
\end{eqnarray}

\section{Direct Integration, Method B}
\label{appndxB}
We here start with the integrating factor $\mu(t)=\exp(\frac{\alpha}{m}(t-t_n))$, which gives the differential equation from Eq.~(\ref{eq:Langevin})
\begin{eqnarray}
m\frac{d(\mu v)}{dt} & = & (f+\beta)\mu\; . 
\end{eqnarray}
The solution to this equation is
\begin{eqnarray}
v(t) & = & \dot{r}(t) \; = \; \frac{1}{\mu(t)}v^n+\frac{1}{\mu(t) m}\int_{t_n}^t\mu(s)(f(s)+\beta(s))\,ds\; , \label{eq:v_of_t_2}
\end{eqnarray}
which for $t=t_{n+1}$ can be written
\begin{eqnarray}
v(t_{n+1}) & = &  \tilde{c}_2\left(v^n+\frac{1}{m}\int_{t_n}^{t_{n+1}}\mu(s) f(s)\,ds\right)
 +\frac{\tilde{d}_{Bv}}{m}\beta_{Bv}^{n+1}\; , \label{eq:v_of_tn_2}
\end{eqnarray}
where $\tilde{c}_2$ is given by Eq.~(\ref{eq:tc2}), and 
\begin{eqnarray}
\tilde{d}_{Bv}\beta_{Bv}^{n+1} & = & \int_{t_n}^{t_{n+1}}e^{\frac{\alpha(s-t_{n+1})}{m}} \beta(s)\,ds\; . \label{eq:tdBvbeta}
\end{eqnarray}
Thus, $\beta_{Bv}^n$ is a Gaussian variable with $\langle\beta_{Bv}^n\rangle=0$ and
\begin{eqnarray}
\tilde{d}_{Bv}^2\langle\beta_{Bv}^n\beta_{Bv}^l\rangle & = & \underbrace{\tilde{c}_1\tilde{c}_3}_{\tilde{d}_{Bv}^2}\,2\alpha\,k_BT\,\Delta{t}\,\delta_{n,l} \; = \;  \tilde{d}_{Bv}^2\,2\alpha\,k_BT\,\Delta{t}\,\delta_{n,l}\; .  \label{eq:tdv}
\end{eqnarray}

Direct integration of Eq~(\ref{eq:v_of_t_2}) gives
\begin{eqnarray}
r(t)-r^n & = & v^n\int_{t_n}^t\frac{1}{\mu(s)}\,ds+\frac{1}{m}\int_{t_n}^t\frac{1}{\mu(t^\prime)}\int_{t_n}^{t^\prime}\mu(s)(f(s)+\beta(s))\,ds\,dt^\prime\\
 & = & \frac{1-e^{-\frac{\alpha}{m}(t-t_n)}}{\alpha(t-t_n)/m}(t-t_n)v^n +\frac{1}{m}\int_{t_n}^t(f(s)+\beta(s))\frac{1-e^{\frac{\alpha}{m}(s-t_{n+1})}}{\alpha/m}\,ds\; .
\end{eqnarray}
At $t=t_{n+1}$ this expression reads
\begin{eqnarray}
r(t_{n+1}) & = & r^n
+\tilde{c}_3\Delta{t}\,v^n +\frac{\Delta{t}}{m}\int_{t_n}^{t_{n+1}}f(s)\frac{1-e^{\frac{\alpha}{m}(s-t_{n+1})}}{\alpha\Delta{t}/m}\,ds+\tilde{d}_r\frac{\Delta{t}}{2m}\beta_r^{n+1}\; , \label{eq:r_of_tn_2}
\end{eqnarray}
where $\tilde{c}_3$ is given by Eq.~(\ref{eq:tc3}), and  $\beta_r^n$ and $\tilde{d}_r$ are given by Eqs.~(\ref{eq:drbeta_r})-(\ref{eq:tdr}).

The crosscorrelation between the two noise terms is
\begin{eqnarray}
\zeta_B & = & \frac{\langle\beta_{Bv}^n\beta_r^l\rangle}{\sqrt{\langle\beta_{Bv}^n\beta_{Bv}^n\rangle\,\langle\beta_r^n\beta_r^n\rangle}} \; = \; \frac{\tilde{c}_3^2}{\sqrt{1-2\tilde{c}_3+\tilde{c}_1\tilde{c}_3}\sqrt{\tilde{c}_1\tilde{c}_3}}\; . \label{eq:Bcorrelation_correct}
\end{eqnarray}

\begin{figure}[t]
\centering
\scalebox{0.5}{\centering \includegraphics[trim={2.5cm 4.0cm 1.0cm 8.0cm},clip]{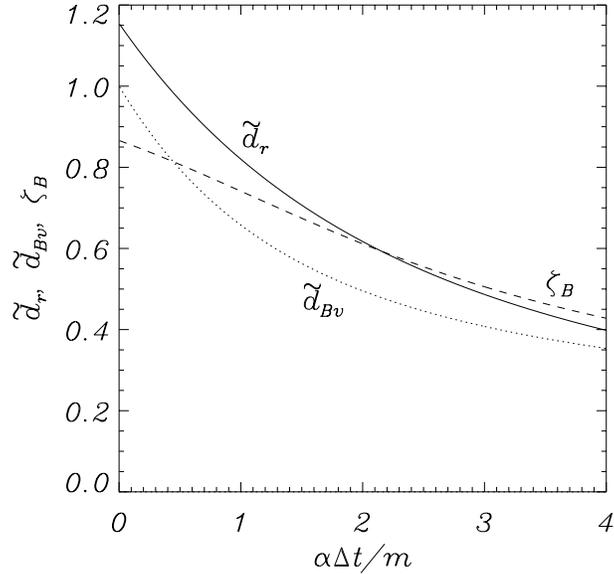}}
\caption{Noise parameters for the direct integration approach of Appendix \ref{appndxB} as a function of reduced time step $\alpha\Delta{t}/m$. Displayed parameters $\tilde{d}_r$, $\tilde{d}_{Bv}$, and $\zeta_B$ are from Eqs.~(\ref{eq:tdr}), (\ref{eq:tdv}), and (\ref{eq:Bcorrelation_correct}), respectively.}
\label{fig_App_B}
\end{figure}
Please see Fig.~\ref{fig_App_B} for the noise parameters $\tilde{d}_r$, $\tilde{d}_{Bv}$, and $\zeta_B$. 

The resulting numerical method appears by making the appropriate trapezoidal approximations to the integrals over the conservative force $f$ in Eqs.~(\ref{eq:r_of_tn_2}) and (\ref{eq:v_of_tn_2}):
\begin{subequations}
\begin{eqnarray}
r^{n+1} & = & r^n+\tilde{c}_3\Delta{t}\left[v^n+\frac{\Delta{t}}{2m}f^{n}\right]+\frac{\Delta{t}}{2m}\tilde{d}_r\beta_r^{n+1}\label{eq:Method_Br}\\
v^{n+1} & = & \tilde{c}_2v^n+\frac{\Delta{t}}{2m}(\tilde{c}_2f^n+f^{n+1})
 +\frac{1}{m}\tilde{d}_{Bv}\beta_{Bv}^{n+1}\; . \label{eq:Method_Bv}
 \end{eqnarray}\label{eq:Method_2}
\end{subequations}

Equation (\ref{eq:Method_2}) can finally be written
\begin{eqnarray}
r^{n+1} & = & 2\tilde{c}_1r^n-\tilde{c}_2r^{n-1}+\frac{\tilde{c}_3\Delta{t}^2}{m}f^n
+\frac{\Delta{t}}{2m}\left(2\tilde{c}_3\tilde{d}_{Bv}\beta_{Bv}^n-\tilde{c_2}\tilde{d}_r\beta_r^n+\tilde{d}_r\beta_r^{n+1}\right)\; , 
\end{eqnarray}
where $\tilde{c}_1$ is given by Eq.~(\ref{eq:tc1}).

\section{Boltzmann distribution in a harmonic potential, $c_3^\prime\neq0$}
\label{appndxC}
We here wish to investigate if the special parameter case of $c_3^\prime\neq0$ (see Eqs.~(\ref{eq:general_rr}) and (\ref{eq:c3prime})) is relevant for the development of statistically correct methods. This special case is made possible by the special velocity parameter case $\gamma_3=0$, as described in Sec.~\ref{sec:general} through Eqs.~(\ref{eq:general_vv}) and (\ref{eq:general_vv_prime}), and comments thereafter.
 
For $f^n=-\kappa r^n=-\Omega_0^2mr^n$, where $\Omega_0=\sqrt{\kappa/m}$ is the natural frequency of the harmonic oscillator with spring constant $\kappa>0$, we adopt the methodology of Ref.~\cite{GJ} to find the autocorrelation $\langle r^nr^n\rangle$ from the linearized version of Eq.~(\ref{eq:general_rr}) (see Eq.~(\ref{eq:Cgeneral_rr_lin}) below). In order to simplify the visual impression of the expressions, we use Eq.~(\ref{eq:Big_noise})
\begin{eqnarray}
{\cal N}^{n+1} & = & \frac{\Delta{t}}{2m}\left(2c_4\beta_v^n-c_5\beta_r^n+c_6\beta_r^{n+1}\right) \nonumber
\end{eqnarray}
and define
\begin{eqnarray}
X & = & 1-\frac{c_3}{c_1}\frac{\Omega_0^2\Delta{t}^2}{2}\\
Y & = & 1+\frac{c_3^\prime}{c_2}\Omega_0^2\Delta{t}^2
\end{eqnarray}
such that the linearized  Eq.~(\ref{eq:general_rr}) can be written
\begin{eqnarray}
r^{n+1} & = & 2c_1Xr^n-c_2Yr^{n-1}+{\cal N}^{n+1}\; . \label{eq:Cgeneral_rr_lin}
\end{eqnarray}

Multiplying Eq.~(\ref{eq:general_rr_lin}) with, respectively, $r^{n-1}$, $r^n$, and $r^{n+1}$,  and then making the statistical averages of the resulting equations, we get
\begin{eqnarray}
\left(\begin{array}{ccc}
1 & -2c_1X & c_2Y \\
0 & 1+c_2Y & -2c_1X \\
c_2Y & -2c_1X & 1\end{array}\right)\left(\begin{array}{c}\langle r^{n-1}r^{n+1}\rangle \\ \langle r^nr^{n+1}\rangle \\ \langle r^nr^n\rangle\end{array}\right) & = & \left(\begin{array}{c}\langle r^{n-1}{\cal N}^{n+1}\rangle \\ \langle r^{n}{\cal N}^{n+1}\rangle \\ \langle r^{n+1}{\cal N}^{n+1}\rangle \end{array}\right)\; , 
\end{eqnarray}
which can also be written
\begin{eqnarray}
\left(\begin{array}{ccc}
1 & -2c_1X & c_2Y \\
0 & 1+c_2Y & -2c_1X \\
0 & -2c_1X & 1+c_2Y\end{array}\right)\left(\begin{array}{c}\langle r^{n-1}r^{n+1}\rangle \\ \langle r^nr^{n+1}\rangle \\ \langle r^nr^n\rangle\end{array}\right) & = & \left(\begin{array}{c}\langle r^{n-1}{\cal N}^{n+1}\rangle \\ \langle r^{n}{\cal N}^{n+1}\rangle \\ \frac{1}{1-c_2Y}\langle r^{n+1}{\cal N}^{n+1}\rangle \end{array}\right)\; . 
\end{eqnarray}
The most desired autocorrelation is then directly given by
\begin{eqnarray}
\langle r^nr^n\rangle & = & \frac{(1+c_2Y)\langle r^{n+1}{\cal N}^{n+1}\rangle+(1-c_2Y)2c_1X\langle r^n{\cal N}^{n+1}\rangle}{(1-c_2Y)\left[(1+c_2Y)^2-(2c_1X)^2\right]} \; = \; \frac{k_BT}{\kappa}\label{eq:Crnrn_1}\\
 & = & \frac{4c_1X\langle r^{n}{\cal N}^{n+1}\rangle +(1+c_2Y)\left(\frac{\Delta{t}}{2m}\right)^2\left[4c_4^2-4c_4c_5\zeta+c_5^2+c_6^2\right]\,2\alpha\,k_BT\,\Delta{t}}{(1-c_2Y)\left[(1+c_2Y)^2-(2c_1X)^2\right]}\\
 & = & \frac{k_BT}{\kappa} \frac{\alpha\Delta{t}}{2m}\Omega_0^2\Delta{t}^2\, \frac{4c_1Xc_6(2c_4\zeta-c_5) +(1+c_2Y)\left[4c_4^2-4c_4c_5\zeta+c_5^2+c_6^2\right]}{(1-c_2Y)\left[(1+c_2Y)^2-(2c_1X)^2\right]}\\
& = & \frac{k_BT}{\kappa} \frac{\alpha\Delta{t}}{2m}\, \frac{4c_1Xc_6(2c_4\zeta-c_5) +(1+c_2Y)\left[4c_4^2-4c_4c_5\zeta+c_5^2+c_6^2\right]} {(1-c_2-c_3^\prime\Omega_0^2\Delta{t}^2)[4c_1(c_3^\prime+c_3)+((c_3^\prime)^2-c_3^2)\Omega_0^2\Delta{t}^2]}\; . 
\end{eqnarray}
This leads to the requirement that one of the following two conditions must be fulfulled
\begin{subequations}
\begin{eqnarray}
c_3^\prime & = & 0\label{eq:cond_C0}\\
c_3^\prime & = & c_3\; . \label{eq:cond_C1}
\end{eqnarray}
\end{subequations}
The condition Eq.~(\ref{eq:cond_C0}) is considered in Sec.~\ref{sec:harmonic}. Thus, we will here consider the condition Eq.~(\ref{eq:cond_C1}).
With that constraint we get
\begin{eqnarray}
\langle r^nr^n\rangle & = &  \frac{k_BT}{\kappa} \frac{\alpha\Delta{t}}{2m}\, \frac{4c_1Xc_6(2c_4\zeta-c_5) +2c_1(2-X)\left[4c_4^2-4c_4c_5\zeta+c_5^2+c_6^2\right]} {(2c_1X-2c_2)4c_12c_3}\; , 
\end{eqnarray}
which leads to the two separable conditions
\begin{eqnarray}
-4c_2c_3 & = & \frac{\alpha\Delta{t}}{2m}\left[4c_4^2-4c_4c_5\zeta+c_5^2+c_6^2\right]\label{eq:cond_C2}\\
2c_1c_3 & = & \frac{\alpha\Delta{t}}{2m}\left[c_4(2\zeta c_1c_6-c_4)-c_1^2c_6^2\right]\; . \label{eq:cond_C3}
\end{eqnarray}
Notice that the one-time-step velocity attenuation factor must have the limit $c_2\rightarrow1$ for $\alpha\Delta{t}/m\rightarrow0$. This implies that $c_1\rightarrow1$ and $c_3\rightarrow\frac{1}{2}$ for $\alpha\Delta{t}/m\rightarrow0$. Thus, the left hand sides of Eqs.~(\ref{eq:cond_C2}) and (\ref{eq:cond_C3}) limit the values $-2$ and $1$, respectively, while the right hand sides limit $0$, unless at least one of the functional parameters $c_4$, $c_6$ or $\zeta$ diverges for $\alpha\Delta{t}/m\rightarrow0$. This does not seem reasonable, and we therefore conclude that the case $c_3^\prime\neq0$ ($\gamma_3=0$) is not leading to a meaningful method that converges correctly for $\alpha\Delta{t}/m\rightarrow0$.

\end{document}